\documentclass[twocolumn,preprintnumbers,amsmath,amssymb]{revtex4}

\usepackage{graphicx}
\usepackage{dcolumn}
\usepackage{bm}

\begin{document}

\title{Quark Gluon Bags as Reggeons}

\author{K. A. Bugaev, V. K. Petrov  and G. M. Zinovjev}
\affiliation{Bogolyubov Institute for Theoretical Physics,
Kiev, Ukraine
}

\date{\today}
\begin{abstract}
The  influence  of the  medium dependent finite width of quark gluon plasma (QGP) bags
on their equation of state is analyzed within an exactly solvable model. 
It is argued that the  large width of the QGP bags  not only explains 
the observed deficit in the number of  hadronic resonances, but also clarifies the reason   why 
the heavy QGP bags   cannot be directly observed  as metastable  states in a hadronic phase. 
The model allows us  to estimate the minimal 
value  of the width of QGP bags  being heavier than 2 GeV 
 from a variety of the lattice QCD data and get  that 
the minimal  resonance width at zero temperature is about 600 MeV,
whereas  the minimal resonance width at  the  Hagedorn temperature is about 2000 MeV.  As shown,
these estimates are almost  insensitive to the number of the elementary degrees of freedom. 
The recent lattice QCD data are analyzed and it is found that besides $\sigma T^4$   term  the lattice QCD pressure 
contains $T$-linear and $T^4 \ln T$ terms in the range 
of temperatures between 240 MeV and 420 MeV. The presence of the last term in the pressure   
bears almost  no effect on the width estimates. 
Our analysis  shows  that  at  high  temperatures the average mass and width of the QGP bags  
behave in accordance with   the upper bound of the Regge trajectory asymptotics (the linear asymptotics), 
whereas at low temperatures they obey  the lower bound 
of   the Regge trajectory asymptotics (the square root one). Since the model explicitly contains 
the Hagedorn mass spectrum,  it allows us to remove  an existing contradiction between the finite number of  
hadronic Regge families and the Hagedorn idea of the exponentially growing  mass spectrum of hadronic bags.

\vspace*{0.25cm} 

\end{abstract}

\maketitle


\section{Introduction}

The concept of strongly interacting  QGP (sQGP),
\cite{Shuryak:sQGP,Shuryak:sQGPb} has created 
a new framework for the  quantum chromodynamic (QCD) 
phenomenology. However, the very strong  argument that sQGP is not a
weakly coupled gas,  but rather a strongly coupled liquid 
\cite{Shuryak:sQGP}, 
does  not  clarify the question of what are  the relevant degrees of freedom to formulate
the proper 
statistical description of  the sQGP.
The idea that the dressed quarks and gluons should form the multiple  binary colored bound states \cite{Shuryak:sQGP} and even multibody bound states \cite{Shuryak:sQGP,MultColor}
does not help much, since in the sQGP the colored objects  must strongly interact 
with the surrounding media and form the colorless clusters  whose  interaction with  each other  is essentially reduced compared to their   constituents.
Such a behavior of  interacting constituents is typical for  the clusters formed by
the molecules in the ordinary liquids 
\cite{Fisher:67, Elliott:06} and by the nucleons in a nuclear liquid 
\cite{Bondorf:95,Bugaev:00, Elliott:06}.
An existence of  colorless clusters in the sQGP  is indirectly supported by the huge quark-antiquark   ``energy" potentials  
found from the lattice QCD (LQCD) simulations \cite{Shuryak:sQGPb,LQCD}, which indicate 
us that  at energy densities right above the deconfinement 
transition there is no separation of valence quarks belonging to the same hadron.  
These huge values of binding energy  indicate that the relevant degrees of freedom 
in the sQCD are  heavy  and large hadrons which nowadays are regarded as the QGP bags.  

The idea that the relevant degrees of freedom in the QGP are the hadronic bags  of any volumes and masses that contain quarks and gluons,   inside  was first formulated in Ref.  \cite{Kapusta:81} and named,  gas of bags model (GBM). This work has unified several  instructive  results obtained earlier: it was shown  \cite{Kapusta:81} that the MIT bag model \cite{MITBagM} leads
to the Hagedorn mass spectrum of bags \cite{Hagedorn:65} and  the phase transition to the QGP  corresponds to a formation of the infinitely large  bag. 
Further development in this direction led to many interesting findings \cite{Bugaev:07,CGreiner:06}.
The most promising of them is an
inclusion of 
 the quark gluon bags surface tension into statistical description  \cite{Bugaev:07} which allows one to  simultaneously 
describe the 1-st and 2-nd order deconfinement phase transition with the cross-over.

The existence of the QGP bags with the Hagedorn mass spectrum  near the transition temperature to hadronic phase is strongly supported by the fact that the Hagedorn 
resonances can naturally explain the extra baryon (antibaryon) \cite{BantiB} and kaon (antikaon) \cite{KantiK} production which was found  in the 5 \%
most central Au-Au collisions at RHIC  \cite{ExtraB}.  
Also it seems that the  QGP bags with the Hagedorn mass spectrum 
can explain the fast  chemical equilibration of hadrons in an expanding fireball 
\cite{KantiK}. Such an equilibration   is a manifestation of the fact that the resonances 
with Hagedorn mass spectrum  are the perfect particle reservoirs and perfect thermostats 
\cite{Moretto:06}.  
Hence our starting point is the quark gluon bags with surface tension model  (QGBSTM) 
\cite{Bugaev:07}, which includes all the  important features of the  previously suggested
statistical models discussed above. 

The question of the relevant degrees of freedom in the sQGP is of  principle importance not only to  formulate the correct statistical description, but also 
to explain 
the various phenomena   which occur in the sQGP  during its thermalization, expansion
and hadronization in the course of high energy nuclear or elementary particle collisions.
We mean not only the direct flow, conical flow or the jet quenching, but the mechanism 
of  thermal and chemical equilibration of the sQGP 
whose investigation  requires the development of nonperturbative methods suited  to describe such  processes.  Unfortunately, the powerful methods suggested in the past were not  developed further 
to exploit  them for  the present  needs of heavy ion phenomenology.
For example, the well known  dual resonance model 
\cite{DualRM} is able  to explain not only  the Hagedorn mass spectrum,  but also the process of thermalization of decaying resonances \cite{DualRM:Thermaliz}.
Hence it could be a good starting point to study the process of thermalization 
of the QGP bags, but  due to the  lack  of  our  knowledge  on the relevant degrees of freedom of  sQGP  it is as yet unclear   to what extent  is it possible to use  the  dual resonance model for this purpose.  

The problem of the relevant degrees of freedom in the sQGP has another important aspect.  
Indeed, the masses and bounds on charge or isospin of  the sQGP constituents 
are intensively discussed in the LQCD community \cite{Karsch:Fluct06},
but at the same time   important characteristics such as their 
mean volume and life-time have not  yet
caught the  necessary attention.  However,  these quantities may put some
new bounds on the spatial and temporal properties of the sQGP \cite{FWM:08} created in 
high energy collisions. If, for instance, the sQGP consists of droplets of finite (mean) size,
then one could naturally resolve the HBT puzzles at RHIC energies \cite{Wong:HBT}.
On the other hand, the short life-time of heavy QGP bags found recently \cite{FWM:08} may not only play an  important role in  all thermodynamic and hydrodynamic phenomena of the sQGP matter mentioned above, but may also explain 
the absence of strangelets  \cite{Strangelets:06} or, more generally,  why the finite  QGP bags cannot be observed
at energy densities typical for hadronic phase \cite{FWM:08} (see below). 
Therefore, an investigation of the  mass and volume distributions and the life-time of the QGP bags and their  consequences for both the experimental observables and 
theoretical  studies   is  very  pertinent.  

The work is organized as follows. 
First  we  thoroughly   discuss  the two conceptual problems of the GBM and its generalizations 
which   are typical for finite systems created in  high energy collisions.  Then in Sect. III  we 
present  the  finite width model (FWM) of QGP bags \cite{FWM:08}  and in Sect. IV  we 
show how  both conceptual  problems can be naturally resolved 
within  the FWM  which is a principally 
new kind of model compared to the GBM generalizations and early attempts  to  derive 
the mass-volume spectrum of the QGP bags \cite{Goren:82,CGreiner:06}. 
Sect. V is devoted to the analysis of  the LQCD pressure and trace anomaly.
These results allow us to 
 estimate the width of the QGP  bags  from the LQCD thermodynamics in Sect. VI. 
In Sect. VII we compare  two regimes of the FWM and show that in the high pressure 
regime the FWM  QGP bags  behave in accordance with   the upper bound  \cite{Trushevsky:77} of  the  asymptotic  behavior of the Regge trajectories for the  mass and width of hadronic resonances,
whereas in the low pressure limit  they obey the lower limit of 
the  asymptotic  behavior of the Regge trajectories  \cite{Trushevsky:77}.   Thus, we   explicitly demonstrate
that the large and/or heavy  QGP bags can be regarded as 
the objects belonging to the Regge trajectories. 
Furthermore  we establish  the close relations between the Hagedorn idea  of exponentially increasing 
hadronic mass spectrum and finite number of Regge trajectories of QGP bags  within the Regge poles method. 
Our conclusions are formulated in Sect. VII.


\vspace*{0.25cm}

\section{Conceptual Problems of GBM}

Despite  the positive  features of  the GBM \cite{Kapusta:81, Gorenstein:81} and its generalizations  
\cite{CGreiner:06,Bugaev:07, Goren:05}, all of  them  face  two conceptual problems. 
The first one can be formulated by asking a very simple question: 'Why are the QGP bags never directly observed  in the experiments?' The routine argument applied to both high energy heavy ion and hadron collisions is that there exists  a phase transition and, hence, the huge energy gap separating the QGP bags from the ordinary (light) hadrons prevents the QGP co-existence at the hadron densities below the phase transition.
The same line of arguments is also valid, if   the strong 
cross-over exists.
The problem, however,  arises from the fact that in the laboratory  experiments we are dealing  with finite systems. From the finite volume  exact analytical solutions of  the constrained statistical multifragmentation model (SMM) \cite{Bondorf:95,Bugaev:00} and  the GBM \cite{Kapusta:81, Gorenstein:81}, found in  
\cite{Bugaev:04a} and \cite{Bugaev:05c,Bugaev:07b}, respectively,  it is known that in  thermally  equilibrated finite system there is
a non-negligible probability of finding   the small and not too heavy QGP bags, say with the mass of 10-15 GeV, 
 even in the hadronic phase. 
Therefore,  for finite volume systems  created in  high energy nuclear or  elementary particle  collisions such QGP bags  could appear as  any other  metastable states in statistical mechanics,
{since in this case  the statistical suppression is just a few orders  of magnitude and not of the order of   the Avogadro number.}

Moreover, the finite volume solution of the GBM  \cite{Bugaev:05c}, in which 
the mean mass of the QGP bag is proportional to its volume,  predicts 
the decay time $\tau_n \approx \frac{V}{\pi n V_0  T}$
for  the collective state $n$ ($ n = 1, 2, 3, \dots$, $V_0 \approx1 $ fm$^3$ 
\cite{FWM:08,Bugaev:05c}) of the  mixed phase having the finite volume
$V$ and temperature $T$.  Therefore, if  a single statistical   state with $n \ge  1$ had a pressure close to zero, its life-time would be determined  by the temperature and the volume.
If, in addition, such a state could  emit the photons or dileptons to reduce its temperature 
without an essential reduction of its volume, it could live for a very  long time compared to the typical  
life-time of heavy hadronic resonances. In particular,  one could think of the strangelets 
\cite{Strangelets:06}, as a  possible example for such a state. 

Consequently, if  such QGP bags  can be created 
 in high energy nuclear  and   in elementary particle collisions  or in some astrophysical phenomena there must be a reason which 
prevents their  direct  experimental  detection. 
 As we will show in the following   there is an inherent property of the strongly 
interacting matter equation of state (EoS) 
which prevents the appearance of such QGP bags 
inside of the  hadronic phase even in finite systems and which  is also responsible for  the instability of  large or heavy  strangelets.

The second conceptual  problem is rooted in   a huge deficit in the  number of observed  hadronic resonances \cite{Bron:04}  with masses above 2.5 GeV predicted by the  Hagedorn model \cite{Hagedorn:65}  and used, so far,  by all other subsequent  models discussed above. 
Thus, there is a paradox  situation with the Hagedorn mass  spectrum: it was predicted for heavy hadrons
which nowadays  must be regarded as QGP bags, but  it can be experimentally  established up to hadronic masses of  about 2.3 GeV  \cite{Bron:04}.
Of course, one could  argue that heavy hadronic resonances cannot be established experimentally 
because both  their   large width  and  very large number of decay channels lead to 
great  difficulties in their identification.  But the point is that, despite the recent efforts of  
Ref. \cite{Blaschke:03},  the influence of   large  width of heavy 
resonances  on their EoS  properties and  the corresponding  experimental consequences 
were  not studied in full. 

The recent  step in this direction was made in  \cite{FWM:08}. 
We  introduced the  finite and medium dependent width into statistical description 
and  studied   its influence on the system's  pressure at vanishing baryonic chemical potential there. 
We argued that  the FWM  mandatory  requires  the inclusion of the width of the QGP bags 
which, on one hand, explains the experimentally observed deficit   of heavy hadronic resonances compared to any of  the previous GBM generalizations, and, on the other hand,
we demonstrated 
that  the new physical effect, {\it  the subthreshold suppression of the QGP bags}  of the FWM,  naturally resolves   the first conceptual problem formulated above.


\section{Basic Ingredients of the  FWM}

The most convenient way to study the phase structure  of  any statistical  model similar to the GBM or QGBSTM  is to use the isobaric partition \cite{Gorenstein:81,Bugaev:07, Bugaev:04a} and find its rightmost singularities. Hence,  we assume that after the Laplace transform  the  FWM grand canonical  partition  $Z(V,T)$ generates the following 
isobaric partition:

\vspace*{-0.5cm}
\begin{align}\label{Zs}
\hspace*{-0.25cm}\hat{Z}(s,T) \equiv \int\limits_0^{\infty}dV\exp(-sV)~Z(V,T) =\frac{1}{ [ s - F(s, T) ] } \,,
\end{align}

\vspace*{-0.2cm}
\noindent
where the function $F(s, T)$ contains the discrete $F_H$ and continuous $F_Q$ mass-volume spectrum 
of the bags 

\vspace*{-0.5cm}
\begin{align}
F(s,T)&\equiv F_H(s,T)+F_Q(s,T) = \sum_{j=1}^n g_j e^{-v_js} \phi(T,m_j) \nonumber 
 %
 \end{align}
 \vspace*{-0.9cm}
\begin{align} 
&+\int\limits_{V_0}^{\infty}dv\hspace*{-0.1cm}\int\limits_{M_0}^{\infty}
 \hspace*{-0.1cm}dm~\rho(m,v)\exp(-sv)\phi(T,m)~.
 \label{FsHQ}
\end{align}

\vspace*{-0.2cm}
\noindent
The   density
of  bags of mass $m_k$, eigen volume $v_k$  and degeneracy $g_k$
is given by  $\phi_k(T) \equiv g_k ~ \phi(T,m_k) $  with 
\vspace*{-0.2cm}
\begin{align} 
\phi_k(T)   & \equiv  \frac{g_k}{2\pi^2} \int\limits_0^{\infty}\hspace*{-0.0cm}p^2dp~
\exp{\textstyle \left[- \frac{(p^2~+~m_k^2)^{1/2}}{T} \right] } =
\nonumber \\
& =  g_k \frac{m_k^2T}{2\pi^2}~{ K}_2 {\textstyle \left( \frac{m_k}{T} \right) }\, .
\end{align}
 
\vspace*{-0.2cm}
\noindent
The mass-volume spectrum $\rho(m,v)$ is the generalization of the exponential mass spectrum 
introduced by Hagedorn \cite{Hagedorn:65}.  Similar to  the GBM and QGBSTM,  the 
FWM bags 
are assumed to have the hard core repulsion of the Van der Waals type which generates the suppression factor proportional to the  exponential of  bag eigen volume $\exp(-sv)$. Since the 
mass-volume spectrum  $\rho(m,v)$ can be written in a form containing the discrete part  $F_H$,
hereafter we will not distinguish the discrete bags 
from the bags of continuous spectrum, 
if their properties are similar. However, we will keep
the sum and integrals in (\ref{FsHQ}) explicitly, since they correspond to different phases of the model.

The first term of Eq.~(\ref{FsHQ}), $F_H$, represents the contribution of a finite number of low-lying
hadron states up to mass $M_0 \approx 2 $ GeV \cite{FWM:08} which correspond to  different  flavors. This function has no $s$-singularities at
any temperature $T$ and can generate a simple pole of the isobaric partition, whereas  the mass-volume spectrum of the bags $F_Q(s,T)$ can be chosen to 
generate an essential  singularity $s_Q (T) \equiv p_Q(T)/T$ which defines  the QGP  pressure $p_Q(T)$  at zero baryonic densities \cite{Gorenstein:81,Goren:82, Bugaev:07}.

It  is known
from the definition of pressure in the grand canonical ensemble
 that in the thermodynamic limit its partition 
  behaves as $Z(V,T)\simeq \exp\left[pV/T \right]$.
An exponentially increasing $Z(V,T)$ generates the rightmost singularity $s^*=p/T$ 
of the function $\hat{Z}(s,T)$ in variable $s$. 
This is because the integral over $V$ in Eq.~(\ref{Zs}) 
diverges at its upper limit for $s < p/T$. 
Therefore, the rightmost singularity 
$s^*$ of $\hat{Z}(s,T)$ gives us the system  pressure:
\begin{align}\label{p-s}
p(T)~=~T~\lim_{V\rightarrow\infty}\frac{\ln Z(V,T)}{V}~=~T~s^*(T)~.
\end{align}
The  singularity 
$s^*$ of $\hat{Z}(s,T)$ (\ref{Zs}) can be calculated from the transcendental
equation \cite{Gorenstein:81,Bugaev:07}  $s^*(T)~=~ F(s^*,T)$.

As long as the number of sorts of  bags, $n$, is finite, the only possible singularities  
of $\hat{Z}(s,T)$ (\ref{Zs}) are simple   poles. 
For example, for the ideal gas ($n = 1; v_1=0; F_Q \equiv 0$ in  Eq.~(\ref{FsHQ}))
$s^*=g_1\phi(T,m_1)$ and thus from  Eq.~(\ref{p-s}) one gets
$p=Tg_1\phi(T,m_1)$ which corresponds to the  grand 
canonical ensemble ideal gas EoS 
for the particles of  mass $m_1$ and degeneracy $g_1$.
However, for  an infinite number of sorts of  bags, i.e. for $F_Q \neq 0$,  there may appear an essential  singularity of
$\hat{Z}(s,T)$ which corresponds to a different phase.  
This  property is  used in the FWM. 

Here we use the parameterization  of  the  spectrum $\rho(m,v)$ introduced in  \cite{FWM:08}. 
It assumes that 
\begin{align}\label{Rfwm}
& \rho (m,v) =   \frac{ \rho_1 (v)  ~N_{\Gamma}}{\Gamma (v) ~m^{a+\frac{3}{2} } }
 \exp{ \textstyle \left[ \frac{m}{T_H}   -   \frac{(m- B v)^2}{2 \Gamma^2 (v)}  \right]  } \,, \\ 
&  \rho_1 (v) = f (T)\, v^{-b}~ \exp{\textstyle \left[  -  \frac{\sigma(T)}{T} \, v^{\varkappa}\right] }\,,
\label{R1fwm}
 \end{align}
where we drop the unimportant dependences. As one can see from (\ref{Rfwm}) the mass spectrum has a Hagedorn like parameterization and  the Gaussian attenuation  around the bag mass
$B v$ ($B$ is the mass density of a bag of a  vanishing width) with the volume dependent  Gaussian  width 
$\Gamma (v)$ or width hereafter. 
We will distinguish it from the true width defined as 
$\Gamma_R = \alpha \, \Gamma (v)$ ($\alpha \equiv 2 \sqrt{2 \ln 2}\,$).
We stress  that the Breit-Wigner attenuation  of  a resonance mass cannot be used in the spectrum 
(\ref{Rfwm}) because in case of finite width it would lead to a divergency of the mass integral in (\ref{FsHQ}) above  $T_H$.

The normalization factor 
obeys the condition

\vspace*{-0.5cm}
\begin{align}\label{Ng}
& N_{\Gamma}^{-1}~ = ~ \int\limits_{M_0}^{\infty}
 \hspace*{-0.1cm} \frac{dm}{\Gamma(v)}
    \exp{\textstyle \left[  -   \frac{(m- B v)^2}{2 \Gamma^2 (v)}  \right] } \,.
 \end{align}

\vspace*{-0.2cm}
\noindent
The constants  $a > 0$ and  $b > 0$ will be specified later.

The present  choice  of mass-volume spectrum  (\ref{Rfwm}) is  a natural  extension   of early attempts  \cite{Goren:82}
to explore the bag volume  as a statistically independent   degree of freedom to derive an internal  pressure 
of large bags. As it will be shown   such a simple parameterization   not only allows  us to resolve both of the conceptual problems discussed  above, but also
 it  gives  us  an exactly solvable model. 
Our further  motivation for  the mass-volume spectrum   (\ref{Rfwm})  is based on two facts: first, for all
known  hadronic resonances
the width and mass are independent characteristics, and, second, in a  dense medium the reaction rates 
may  change and   may  lead  to the medium dependence of the resonance width. 
In order to take both of them into  account   we   introduced the volume dependence into the Gaussian width
and  we  came to conclusion  that the proper  characteristic  to indicate  the impact  of  a medium  is not the resonance width, but  the average resonance width. 
The latter allows us to get simultaneously  the  mass and temperature dependences of the resonance width 
averaged with respect to the resonance volume (see Eq. (8)). 
Moreover, as will be shown later such a parameterization of  the mass-volume spectrum leads  not only  to a single 
choice for the Gaussian width volume dependence, but also allows us to introduce  the concept of  Regge trajectories 
for the averaged quantities of the QGP bags.

The volume spectrum in  (\ref{R1fwm}) contains the surface free energy (${\varkappa} = 2/3$) with the $T$-dependent 
surface tension which is parameterized as 
$\sigma(T) = \sigma_0 \cdot
\left[ \frac{ T_{c}   - T }{T_{c}} \right]^{2k + 1} $  ($k =0, 1, 2,...$) \cite{Bugaev:07, Bugaev:04b},
where  $ \sigma_0 > 0 $ can be a smooth function of temperature.  For $T$ being not larger than   the tricritical temperature $T_{c}$ such a parameterization  is justified by the usual  cluster models 
like the FDM \cite{Fisher:67,Elliott:06} and SMM \cite{Bondorf:95,Bugaev:00,Reuter:01}, whereas 
the general consideration  for any  $T$   can be driven  by  the surface partitions of the Hills and Dales model 
\cite{Bugaev:04b}. In Ref.  \cite{Bugaev:07} it  was  argued  that at low baryonic densities 
the first order deconfinement phase transition degenerates into a cross-over just because of 
negative surface tension coefficient for $ T > T_{c} $. The other  consequences of the present 
surface tension  parameterization and the discussion of the absence of the curvature free energy in 
(\ref{R1fwm}) can be found in Refs. \cite{Bugaev:07, Complement}.

The spectrum (\ref{Rfwm}) has a simple form, but is rather general since both the width $\Gamma (v)$ and the bag mass density $B$ can be medium dependent. It clearly reflects the fact 
that the QGP bags are similar to   the ordinary  quasiparticles with the medium dependent characteristics (life-time, most probable values of  mass and volume). Now we are ready to  derive the infinite bag pressure 
for two choices of the width: the volume independent width $\Gamma(v) \equiv \Gamma_0$ and 
the volume dependent width $\Gamma(v) \equiv \Gamma_1 = \gamma v^\frac{1}{2}$. 
As will be seen  below the latter resolves   both of the conceptual problems discussed earlier, whereas the former  parameterization  is used for a comparison.


\section{Analysis of the  FWM spectrum}

First we note that for large bag volumes ($v \gg M_0/B > 0$) the factor (\ref{Ng})  can be
found as  $N_\Gamma \approx 1/\sqrt{2 \pi} $.   Similarly, one can show that  for heavy free bags  ($m \gg B V_0$, $V_0 \approx 1$ fm$^3$ \cite{FWM:08},
ignoring the  hard core repulsion and thermostate)
\vspace*{-0.05cm}
\begin{align}\label{Rm}
& \rho(m)  ~ \equiv   \int\limits_{V_0}^{\infty}\hspace*{-0.1cm} dv\,\rho(m,v) ~\approx ~
\frac{  \rho_1 (\frac{m}{B}) }{B ~m^{a+\frac{3}{2} } }
\exp{ \textstyle \left[ \frac{m}{T_H}     \right]  } \,.
\end{align}

\vspace*{-0.1cm}
\noindent
It originates in   the fact that  for heavy bags the 
Gaussian  in (\ref{Rfwm}) acts like a Dirac $\delta$-function for
either choice of $\Gamma_0$ or $\Gamma_1$. 
Thus, the Hagedorn form of  (\ref{Rm}) has a clear physical meaning and, hence, it  gives an additional argument in favor of the FWM. Also it gives an upper bound for the 
volume dependence of $\Gamma(v)$: the Hagedorn-like mass spectrum (\ref{Rm}) can be derived, if for large $v$ the width  $\Gamma$ increases slower than $v^{(1 - \varkappa/2)}= v^{2/3}$. 

Similarly to (\ref{Rm}), one can estimate the width of heavy free bags  averaged over bag volumes and get  $ \overline{\Gamma(v) } \approx  \Gamma(m/B) $.
Thus, 
for  $\Gamma_1(v)$ the mass spectrum of heavy free QGP bags 
must be the Hagedorn-like one with the property that  heavy resonances have to develop   the large  mean width $ \Gamma_1(m/B) = \gamma \sqrt{m/B}$ and, hence,  they 
are hardly   observable. 
Applying these arguments to the strangelets,
we conclude  that, if their mean volume is a few cubic fermis or larger, they  should survive  a  very short time,
which  is similar to the results of  Ref. \cite{Strangelets:06}  predicting  an instability of such strangelets.

Note also that such a mean width is essentially different from both the linear mass dependence of string models  \cite{StringW} and from an  exponential  form  of the nonlocal field theoretical models \cite{NLFTM}.
Nevertheless, as we demonstrate   while discussing the Regge trajectories 
(\ref{aUP}) and (\ref{alphaHT}) the mean width $ \Gamma_1(m/B) $ leads to the linear 
Regge trajectory of  heavy free QGP bags for large values of the invariant mass squared.

\begin{figure}
\includegraphics[width=84mm,height=70mm]{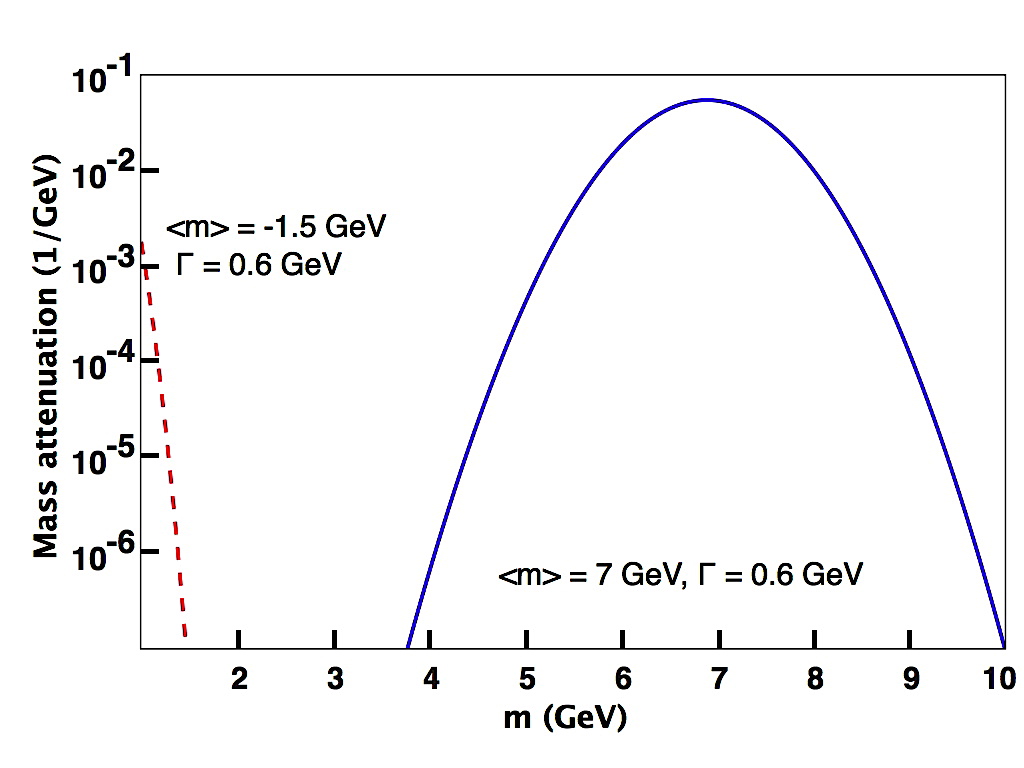}
\caption{[Color online] 
The  resulting mass attenuation $\frac{N_\Gamma\, M_0^a}{\Gamma (v) \, m^a} \exp{ \textstyle \left[  -   \frac{(m- \langle m \rangle)^2}{2 \Gamma^2 (v)}  \right]  }$
 as the function of  bag mass  at the  fixed bag volume for positive and negative values
of the most probable bag mass. The solid  (dashed) curve corresponds to $ \langle m \rangle = 7$ GeV 
($ \langle m \rangle = - 1.5$ GeV). Both curves are shown for the same width $\Gamma (v) = 600$ MeV and $a= 2$.
}
\label{figure1}
\end{figure}

Next we  calculate  $F_Q(s,T)$ (\ref{FsHQ}) for the  spectrum (\ref{Rfwm}) performing the mass integration. There are, however, two distinct  possibilities, depending on the sign of the most probable mass:

\vspace*{-0.2cm}
\begin{align}\label{Mprob}
& \langle m \rangle ~ \equiv ~  B v + \Gamma^2 (v) \beta\,,\quad {\rm with} 
\quad \beta \equiv  T_H^{-1} - T^{-1} \,. 
\end{align}
 %
 %
\noindent
If {\boldmath 
$ \langle m \rangle > 0$} for $v \gg V_0$,  one can use the saddle point method
for mass integration to  find  the function~$F_Q (s,T)$

\vspace*{-0.2cm}
\begin{align}\label{FQposM}
&  F_Q^+ (s,T)   \approx \left[  \frac{T}{2\pi} \right]^{\frac{3}{2} }
\int\limits_{V_0}^{\infty}dv ~ \frac{ \rho_1(v) }{\langle m \rangle^a} ~\exp{\textstyle \left[  \frac{(p^+  - sT )v}{T}  \right]} \, 
\end{align}

\vspace*{-0.1cm}
\noindent
and the pressure of large  bags 
\begin{align}\label{PposMn}
p^+ \equiv T \left[ \beta B + \frac{\Gamma^2 (v)}{2 v} \beta^2 \right] \,. 
\end{align}
To get  (\ref{FQposM}) one has to use  in (\ref{FsHQ}) an asymptotic form  of the $K_2$-function  $\phi(T,m)\simeq (mT/2\pi)^{3/2}\exp(-m/T)$ for $m\gg~T$, 
{collect all terms with $m$ in the exponential, get a full square for $(m -  \langle m \rangle)$ 
and make the Gaussian integration.  The resulting mass attenuation of the obtained spectrum 
$\frac{N_\Gamma\, M_0^a}{\Gamma (v) \, m^a} \exp{ \textstyle \left[  -   \frac{(m- \langle m \rangle)^2}{2 \Gamma^2 (v)}  \right]  }$
at the  fixed bag volume 
is shown in Fig. 1 as  the solid curve for the  typical range of parameters ($a = 2$).
}

Since for $s  <  s_Q^*(T) \equiv p^+(v\rightarrow \infty)/T $ the integral  (\ref{FQposM}) diverges at  its upper limit, 
the  partition (\ref{Zs}) has  an essential singularity that  corresponds to
the QGP pressure of  an  infinite  large   bag.  One concludes that
the width  $\Gamma$ cannot grow faster than $v^{1/2}$ for $v\rightarrow \infty$, otherwise $p^+(v\rightarrow \infty) \rightarrow \infty $ and  $F_Q^+ (s,T)$ diverges for any $s$.
Thus,  for {\boldmath $ \langle m \rangle > 0$} the phase structure of the FWM  with  $\Gamma (v) = \Gamma_1 (v)$ is similar to the QGBSTM \cite{Bugaev:07}.

The volume spectrum of bags  $F_Q^+ (s,T)$ (\ref{FQposM}) is of general nature  and, in contrast  with 
the one suggested in \cite{Goren:82},  has a clear physical meaning. One can also see that  two general 
origins of the bulk part of  bag's  free energy

\vspace*{-0.2cm}
\begin{align}\label{PposM}
&  
- p^+ v = - T \left[ \beta\, \langle m \rangle - \frac{1}{2 } \,\Gamma^2 (v) \beta^2 \right]
\end{align}

\vspace*{-0.1cm}
\noindent
are the bag's  most  probable mass  and its width. Choosing different $T$ dependent functions $\langle m \rangle$ and $\Gamma^2 (v)$, one obtains  different equations of state. 

Comparing the $v$ power of the exponential prefactor in (\ref{FQposM}) to  the continuous 
volume spectrum of bags of the QGBSTM \cite{Bugaev:07}, we find that 
$a + b \equiv \tau \le  2$.

It is  possible to use the spectrum (\ref{FQposM}) not only for infinite system volumes, but also for 
finite volumes $V \gg V_0$. In this case the upper limit of integration should be replaced by finite  $V$ 
(see Refs. \cite{Bugaev:05c, Bugaev:07b} for details).  This will change the singularities of partition 
 (\ref{Zs}) to a set of simple poles  $ s_n^*(T)$ in the complex $s$-plane which are  defined by the same equation as for 
 $V \rightarrow \infty$.  Similarly to the finite $V$ solution of the GBM 
\cite{Bugaev:05c, Bugaev:07b},  it can be shown that for finite $T$ the FWM  simple poles may have  small positive or even negative real part which would lead to a non-negligible contribution of the QGP bags into the total spectrum  $F(s,T)$  (\ref{FsHQ}).
In other words, if the spectrum (\ref{FQposM}), was the only volume spectrum of the QGP bags, then there would exist a finite (non-negligible) probability to find  heavy QGP bags ($m \gg M_0$)  in finite systems  at low temperatures  $T \ll T_c$.  
Therefore, using the results of   the finite volume GBM and SMM,  we  conclude that the spectrum 
(\ref{FQposM}) itself
cannot   explain  the absence of  the QGP bags at  $T \ll T_c$  and, hence, an alternative explanation of this fact is required. 

Such an explanation corresponds to the case {\boldmath$ \langle m \rangle \le 0 $} for $v \gg V_0$.
From (\ref{Mprob}) one can see that  
for the volume dependent width $\Gamma (v) = \Gamma_1 (v) $ the most probable mass $ \langle m \rangle $ inevitably becomes negative at low $T$, if $0 < B < \infty$. 
 In this case the maximum of the Gaussian  mass distribution is located at  resonance masses 
$m = \langle m \rangle \le 0 $. This is true for any argument of the $K_2$-function
 in   $F_Q(s,T)$  (\ref{FsHQ}). 
Since the lower limit of  mass integration  $M_0$ lies above $ \langle m \rangle $, then  only the tail  of  
the  Gaussian mass distribution  may contribute  into $F_Q(s,T)$. 
A  thorough  inspection of the integrand in $F_Q(s,T)$  shows 
(see the dashed curve in Fig. 1 for $a=2$)
that above $M_0$  it is strongly decreasing function of resonance mass and, hence, only the vicinity
of  the lower limit of  mass integration  $M_0$ sizably  contributes into $F_Q(s,T)$.
Applying 
the steepest descent method and  the $K_2$-asymptotic form   for $M_0 T^{-1} \gg 1$
one obtains

\vspace*{-0.2cm}
\begin{align}\label{FQnegM}
\hspace*{-0.3cm}F_Q^-(s,T) \hspace*{-0.05cm} \approx  \hspace*{-0.075cm} \left[  \frac{T}{2\pi} \right]^{\hspace*{-0.05cm}\frac{3}{2} } 
\hspace*{-0.15cm}
\int\limits_{V_0}^{\infty} \hspace*{-0.15cm} dv  \frac{ \rho_1(v) N_{\Gamma}\, \Gamma (v)\, \exp{\textstyle \left[  \frac{(p^-  - sT )v}{T}  \right]}
}{M_0^a\, [M_0 - \langle m \rangle  + a \, \Gamma^{2} (v)/ M_0 ]} \hspace*{-0.3cm}
\end{align}

\vspace*{-0.1cm}
\noindent
with the formal expression for the pressure of  QGP bag 

\vspace*{-0.15cm}
\begin{align}\label{pnegM}
p^-\big|_{v \gg V_0}  = {\textstyle  \frac{T}{v} \left[  \beta M_0 -  \frac{(M_0 - Bv)^2}{2\, \Gamma^{2} (v)}  
 \right] }\,.
\end{align}

\vspace*{-0.05cm}
\noindent
We would like to stress that  the last result requires $B>~0$ and it cannot be obtained for a weaker $v$-growth than  $\Gamma(v) = \Gamma_1(v)$. 
Indeed, if $B<0$, then the normalization factor (\ref{Ng}) would not be $1/\sqrt{2 \pi}$, but would become 
$N_{\Gamma} \approx   [M_0 - \langle m \rangle]\, \Gamma^{-1} (v) \exp{\textstyle 
\left[    \frac{(M_0 - Bv)^2}{2\, \Gamma^{2} (v)}  \right]} $ and, thus,  it would cancel 
the leading  term in  pressure (\ref{pnegM}). Note, however, that  the  inequality 
{\boldmath$ \langle m \rangle \le  0 $} for all $v \gg V_0$ with positive $B$ and 
finite $p^-(v \rightarrow \infty)$ is possible
for  $\Gamma(v) = \Gamma_1(v)$ only. In this case the pressure of an infinite bag is
%
\begin{align}\label{pnegM2}
p^-(v \rightarrow \infty)   =  {\textstyle  - T\frac{ B^2}{2\, \gamma^2 } }  \,.
\end{align}

Also it is necessary to point out  that the only width $\Gamma(v) = \Gamma_1(v)$ 
does not lead to any divergency  in the bag pressure in thermodynamic limit. 
This is clearly seen from Eqs. (\ref{PposMn}) and (\ref{pnegM}) since the multiplier 
$\Gamma^2(v)$ stands in the numerator  of the pressure (\ref{PposMn}), whereas 
in the pressure (\ref{pnegM}) it appears in the denominator.  Thus, if one chooses 
the different $v$-dependence  for the  width, then either $p^+$ or $p^-$  would 
diverge for the bag of  infinite  size.

The new outcome of this case with $B>0$ is that for $T < T_H$ the spectrum 
(\ref{FQnegM}) contains the lightest QGP bags having the  smallest volume since 
every term in the pressure (\ref{pnegM}) is negative.  The finite volume of the system is no longer 
important   because only  the  smallest bags survive in (\ref{FQnegM}).
Moreover, if such bags are created, they would have mass about  $M_0$ and
the width about $\Gamma_1(V_0)$, and, hence, they would not be distinguishable 
from the usual low-mass hadrons. 
Thus, the case {\boldmath$ \langle m \rangle \le 0 $} with 
$B>0$ leads to the {\it subthreshold suppression of the QGP bags} at low temperatures,
since their most probable mass is below the mass threshold  $M_0$ of the spectrum $F_Q(s,T)$.   Note that such an effect cannot be derived within  any of  the GBM-kind models  proposed earlier.
The negative values of  $ \langle m \rangle $ that appeared in the 
expressions above 
serve as an indicator of a different  physical
situation comparing to $ \langle m \rangle > 0$, but    have no physical meaning since 
$ \langle m \rangle \le 0 $ does not enter   the main physical observable  $p^- $.

\section{Comparison with LQCD results}

The obtained results give us an instructive  opportunity to make a bridge between the
particle phenomenology, some experimental facts and the LQCD. 
For instance, if the most probable mass of the QGP bags is known along with the QGP pressure, one can estimate the width of these  bags directly from Eqs. (\ref{PposM})
and  (\ref{pnegM}). 
To demonstrate the new possibilities 
let's now consider   several examples of the QGP  EoS  and relate them  to the above results. First, we study the possibility of  getting   the MIT bag model pressure 
 $p_{bag} \equiv \sigma T^4 - B_{bag} $  \cite{MITBagM}  by  the  stable QGP bags, i.e. 
$\Gamma (v) \equiv 0$. Equating the pressures $p^+$ and $p_{bag}$, one finds that 
the Hagedorn temperature is  related to a bag constant 
$B_{bag} \equiv \sigma T^4_H$.  Then the mass density of such bags  $\frac{\langle m \rangle}{v} $ is identical to 
\begin{equation}\label{EqXVI}
B  =  \sigma T_H (T  + T_H)(T^2 + T_H^2)\,,
\end{equation}
and, hence, it is
always positive. Thus, the MIT bag model  EoS  can be easily obtained within 
the FWM, but, as was discussed earlier, such bags should  be  observable.

Second,  we consider the stable bags, $\Gamma (v) \equiv 0$, but without the Hagedorn 
spectrum, i.e. $T_H \rightarrow \infty$. 
Matching $p^+ = - B $ and $p_{bag}$,  we find that at low temperatures 
the bag mass density  $\frac{\langle m \rangle}{v} = B$ is positive, whereas for 
high $T$ the mass density cannot be positive and, hence,  
one cannot reproduce $p_{bag}$ because in this case $B \le 0$ and  the resulting pressure is not $p^-$ (\ref{pnegM}),
but rather  a zero, as seen from (\ref{FQnegM}), (\ref{pnegM}) and   $N_{\Gamma}$ expression for the limit $\Gamma (v) \rightarrow 0$.

One can try to reproduce $p_{bag}$ with the finite $T$ dependent  width $\Gamma (v) = 2 \, \sigma T^5 v$ for  $T_H \rightarrow \infty$. Then one can get  $p_{bag}$ from 
$p^+$, but only for low temperatures obeying the inequality 
$\frac{\langle m \rangle}{v} = B_{bag} - 2\, \sigma T^4 > 0$. Thus,  the last two 
examples  show us that without the Hagedorn mass spectrum one can not  get 
the MIT bag model pressure. 

The FWM is a phenomenological model including  two independent functions, $B$ and 
$\gamma$,  which 
parameterize    the QGP bag pressure and  require  additional information as an input. 
However, the FWM  provides us with some general results. 
Equating $p^+$ and $p^-(v \rightarrow \infty)$, one can find the transition width coefficient  and  pressure  as 

\vspace*{-0.15cm}
\begin{align}\label{pTR}
\gamma^2_\pm = - \frac{B}{ \beta}, \quad 
p^\pm =   \frac{B T \beta} {2 }\,,
\end{align}

\vspace*{-0.05cm}
\noindent
and one can easily get  that this  transition, indeed, corresponds to 
$\langle m \rangle = 0$.  Thus, although  both expressions for pressure were obtained by 
different methods they match at the correct value of the most probable mass. 
Since $B > 0$ and $\gamma^2_\pm >0$ it follows that such a   transition must occur  at some
temperature $T_\pm = c_\pm \,T_H$ which is below $T_H$, i.e. $ 0 < c_\pm < 1$. 

Another general conclusion concerns   the temperature dependence of the QGP pressure in the limit 
$T \rightarrow 0$. For nonvanishing $\gamma_0 \equiv \gamma (T=0) > 0$  there are, however, two possibilities. The first one corresponds to finite  $B_0 \equiv B (T=0) >0$ values. 
Then from (\ref{pnegM2}) one concludes that  in the limit  $T \rightarrow 0$ the QGP pressure 
linearly depends on temperature $p^-(v \rightarrow \infty) \rightarrow - T\frac{ B_0^2}{2\, \gamma_0^2 } $. 
The second possibility corresponds to the divergent behavior of $B \rightarrow \frac{g_0}{T^D}$ (with $D > 0$)
provided that $\frac{\langle m  \rangle}{v} < 0$ for $ v \rightarrow \infty$.
The latter requires that $D \le 1$ for finite $\gamma_0$. 
In this case at  $T \rightarrow 0$ the QGP pressure should  behave  as 
$p^-(v \rightarrow \infty) \rightarrow - \frac{ g_0^2}{2\, \gamma_0^2 } \, T^{1 - 2D} $. 
Note that either of these possibilities  is a manifestation of the nonperturbative effect since in the limit $\gamma = 0$
they cannot be obtained.

\begin{figure}
\includegraphics[width=84mm,height=63mm]{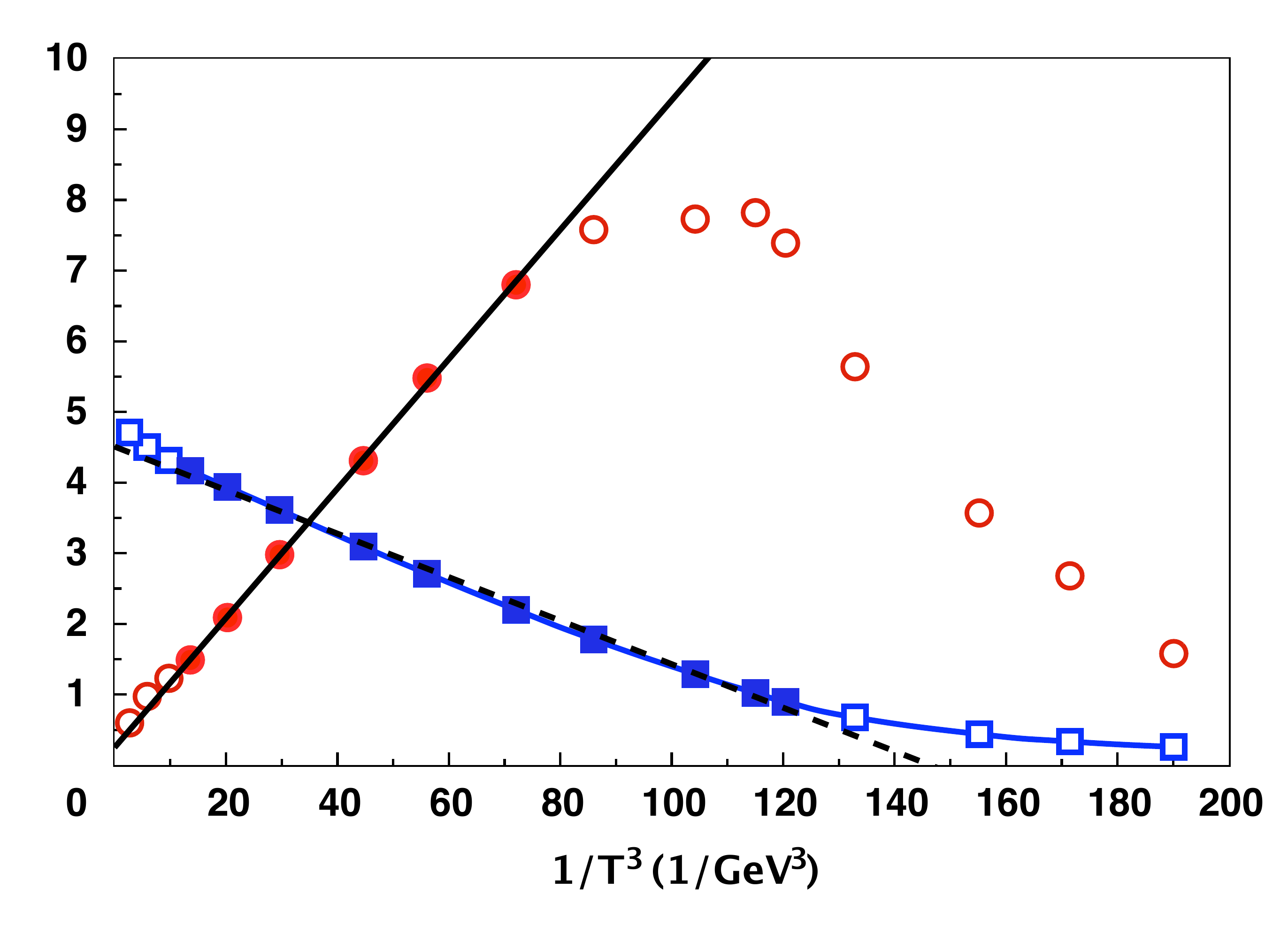}
\caption{[Color online] 
LQCD data for trace anomaly 
(circles) and  pressure per $T^4$ (squares)
as the functions of $T^{-3}$. Straight lines represent the fit of the filled symbols. 
See details in the text. The curve connecting the squares is to guide the eyes. 
}
\label{figure2}
\end{figure}

In Refs.   \cite{LQCD:0,GorMog,LQCD:01} it was  reported that the LQCD data 
exhibit the first of these possibilities. 
The corresponding EoS of the QGP has an additional  linear temperature dependence
\begin{equation}\label{Plin}
p_{lin} = \sigma T^4 - A_1 T + A_0 
\end{equation}
with $ A_1 > 0$, $ A_0 
\ge  0$.
However, the recent analysis \cite{Rob:07} of new  LQCD data \cite{LQCD:96} demonstrated not a linear, but the quadratic $T$-dependence of the trace anomaly and pressure
in the range  of temperatures between about $1.1 T_c$ and  $4 T_c$. 
Therefore, to clarify the question of an additional $T$-dependence of the LQCD pressure we 
analyzed  the old LQCD data \cite{LQCD:1,LQCD:2}, in which the finite size effects are accounted for,  and  also we 
fitted the newest LQCD data that are
found for the almost physical quark masses
\cite{LQCD:3}. 
The fit  of  $p \, T^{-4}$  from  \cite{LQCD:3} 
as function of $T^{-3}$ shown in Fig. 2 by dashed line clearly demonstrates 
the linear  $T^{-3}$ dependence  $p \, T^{-4} = a_0 + a_1\, T^{-3} $ with $a_0 \approx 4.5094 $
and  $a_1 \approx - 0.0304$ GeV$^3$ for  ten data points in the range $T \in [202.5; 419.09]$ MeV. 

The linear $T$ dependence of pressure is rooted  in the behavior of the trace anomaly $\delta = (\varepsilon - 3 \,p) T^{-4} $   (here $\varepsilon$ denotes the energy density). Indeed, plotting $\delta$ as the function of $T^{-3}$ (see circles in Fig. 2)  we found three different types of behavior. 
As one can see from Fig. 2  up to  $T^{-3} \approx 72.056$ GeV$^{-3}$ (for $T \ge  240.31$ MeV) the function
$\delta$ grows nearly linearly, and for  $T^{-3} \ge 120.43$ GeV$^{-3}$ (or $T \le 202.5$ MeV)
it decreases nearly linearly, whereas in between these values of $T^{-3}$ the function
$\delta  $ remains almost constant. 
The analysis shows that six LQCD data points of the function $\delta$ which belong  to  the range $T^{-3} \in [13.585; 72.056]$ GeV$^{-3}$   are, indeed, described by 
$\delta = \tilde{a}_0 + \tilde{a}_1\, T^{-3} $ with $\tilde{a}_0 \approx 0.2514 $
and  $\tilde{a}_1 \approx 0.0916$ GeV$^3$ and 
 $\chi^2/d.o.f. \approx 0.063$, i.e. with extremely 
high accuracy.  
The linear  $T^{-3}$-dependence of   $p \, T^{-4}$  is observed in a slightly  wider range of  $T^{-3}$
because of the approximately constant behavior of  the $\delta$ function at the moderate  values of  $T^{-3}$, but  with  lower  quality   fit which,
however, is comparable with that one of Ref.  \cite{GorMog}.

The reason for  lower quality   of the pressure fit    can be seen  from  its relation to  the lattice  trace anomaly

\vspace*{-0.35cm}
\begin{align}\label{pT4}
\hspace*{-0.15cm}
\frac{p_{fit}}{T^4} - \frac{p_{0}}{T_0^4} &=   \int\limits_{T_0}^T \hspace*{-0.0cm} d\,T \, \frac{\delta}{T}  \nonumber \\
&=  \tilde{a}_0 \ln\left[ \frac{T}{T_0} \right] - \frac{\tilde{a}_1}{3}\, \left[ T^{-3} -  T_0^{-3} \right]\,,  
\end{align}
%
  
\noindent
and, hence, one gets   $a_1 = - \frac{\tilde{a}_1}{3} $,  which is well supported by the LQCD data.
The last equality in (\ref{pT4}) is obtained from the linear fit of  $\delta$ and, hence,
$T_0$ and $p_0 \equiv p_{fit} (T_0)$ are the constants of integration. 

Eq. (\ref{pT4}) shows that for the temperatures between $240.31$ and  $419.09$ MeV
the LQCD pressure \cite{LQCD:3}  does not have a constant term, i.e. $A_0 =0$ for 
$p_{lin}$, but
there exist higher order corrections ($T^5$ and higher) to a  pressure.
They  are small in this range of temperatures since $\tilde{a}_0 \ll 1$, but, in principle,  can be taken into account to improve the quality of 
the linear fit of  $p \, T^{-4}$ function found in \cite{GorMog}. 
However, our main point is that either rough or refined analysis of the  modern LQCD data 
strongly suggests an existence of the linear $T$-dependent term in the LQCD pressure 
for $T \in [ 240.31; 419.09]$ MeV.

Since neither the nonrelativistic hadron gas with the hard core repulsion represented by  $F_H(s,T)$ in (\ref{FsHQ}) nor its relativistic analog 
analyzed in \cite{Bugaev:08NPA} can generate the linear $T$ dependence 
of pressure, it is possible that such a dependence is an inherent property of 
the LQCD data.  Assuming this fact, one obtains  that at low $T$ 
the LQCD pressure of the QGP phase  should behave as $p_{LQCD} (T \rightarrow 0) \rightarrow - |a_1| T$. 
Comparing the linear $T$  dependence  of  the LQCD pressure  with the FWM pressure at low temperatures  
(\ref{pnegM2}), we 
conclude  that the present model  with nonzero width coefficient  correctly   grasps  
the nonperturbative features of the QGP EoS  and  we consider it  as  one  of   the strongest arguments in favor of the FWM.

\section{Width Estimate} 


Such a   behavior of the LQCD  pressure   allows us to roughly  estimate 
the width  $\Gamma_1 (V_0)$ and to study the possible restrictions on thermodynamic functions. 
Since the FWM pressure depends on two functions,  in order 
to find them it is necessary to  know the form of the QGP pressure   in the  hadronic phase.  
Unfortunately the present  LQCD data do not provide 
us with such a detailed information and, hence, some additional 
assumptions are inevitable.  
Consider first the pressure (\ref{Plin}) with $A_0 = 0$ \cite{GorMog}  for nonvanishing 
$B_0$ and $\gamma_0$. 
$T_H$  is uniquely  fixed to be a positive solution of equation 
$A_0 = A_1 T_H - \sigma T_H^4$.
Matching $p_{lin}$  with $p^- (v \rightarrow \infty) = - T \frac{B^2}{2\, \gamma^2}   $ for $T \le c_\pm\, T_H$
we can determine  $B/\gamma$ ratio in this region of temperatures. 
On the other hand, equating  $p_{lin}$ and $p^+$, one obtains the width coefficient  for $T \ge c_\pm\, T_H$
\begin{align}\label{gammaI}
\gamma^2 = 2 \, \beta^{-1} [ \sigma T_H T (T^2 + TT_H + T_H^2) - B(T) ] \,. 
\end{align}
To have 
a positive finite width for all $T \ge c_\pm\, T_H$, it is necessary that  $(T-T_H)$ is a divisor 
of the difference staying  in the square brackets of Eq.  (\ref{gammaI}). Then the  simplest possibility is to suppose that 
\begin{align}\label{BI}
B(T) =  \sigma T_H^2  (T^2 + TT_H + T_H^2) 
\end{align}

\vspace*{-0.05cm}
\noindent
for any $T$.  
Evidently, $B(T)$ in (\ref{BI}) is positive and does not vanish at $T= 0$. 
In addition to a simplicity  another  advantage of such a choice is that (\ref{BI}) does not require 
any new constant or  any new function  which is not involved
in (\ref{gammaI}).
Moreover, comparing ansatz (\ref{BI}) with the mass density (\ref{EqXVI}) obtained  for the pure MIT bag model pressure, one can see  that they differ only  by a term  $\sigma T_H T^3$ 
which at  low   $T \le 0.5 \, T_H$  is a negligible  correction to (\ref{BI}). 
Therefore, 
for low temperatures the ansatz (\ref{BI})  looks  quite reasonable because in this region it corresponds to the mass density of the most popular EoS  of  modern  QCD phenomenology.

 It follows from  (\ref{BI})
 that $ \gamma_0^2 = B_0^2 / (2 A_1) = T_H B_0 / 2 = \sigma T_H^5 /2$ for $T= 0$  and that $\gamma^2 = 2 \, T B(T) $ for $T \ge c_\pm  T_H$. 
As an example let's  consider   the true width for the SU(3) color group with two flavors analyzed in \cite{GorMog} (model B in Table I) for $T_c =  200$ MeV. 
Since for $A_0 = 0$  it is found $A_1 \approx (1.5 T_c)^3$ 
and on the other hand the FWM requires  $A_1 = \sigma T_H^3 $,  then 
one obtains $T_H \approx  0.94 T_c$ for $\sigma = \frac{37}{90} \, \pi^2$. 
Thus,  
the true width for the SU(3) color group with two flavors  is
$\Gamma_R (V_0, T=0) \approx 1.22\,  V_0^{\frac{1}{2}} \, T_c^{\frac{5}{2}} \alpha \approx 587 $ MeV and
$\Gamma_R (V_0, T=T_H)   = \sqrt{12}\,\Gamma_R (V_0, T=0) \approx 2034$ MeV. 
These estimates clearly demonstrate  that there is no way to detect the decays of such
short  living  QGP bags even,  if they are  allowed  by the subthreshold suppression. 
The sensitivity of these results to  $T_c$ value for models A and B is given in  Table I,
which supports our main conclusion for the short life-time of the QGP bags. 

The model C in  Table I corresponds to  $p_{lin}$ pressure, but with $A_0 = 0$ and $A_1 = \frac{\tilde{a}_1}{3}$ obtained from the fit of the function $\delta $. 
As one can see from   Table I the minimal width of the QGP bags  found for the same value of 
the 
transition temperature $T_c$  practically 
does not depend on the number of the QGP  degrees of freedom. 
Such a property is an additional  argument in favor of the ansatz  (\ref{BI}).

\begin{table}[h!] \label{VITable}
\caption{The values of the resonance width for different models. Model A corresponds to 
the   pure gluodynamics for the  $SU(2)_C$ color group \cite{LQCD:1}.
Model B describes the  $SU(3)_C$ color group  LQCD  data with two  quark flavors \cite{LQCD:2} and Model C 
corresponds to the   LQCD of $SU(3)_C$ color group  with three quark flavors \cite{LQCD:3}. 
}

\begin{tabular}{cccc}
\hline
 Model  & \quad $T_c$  &  $  \quad \Gamma_R (V_0, 0) $   &   \quad $ \Gamma_R (V_0, T_H) $ \\ 
Ref. &   \quad (MeV) & (MeV)  & (MeV) \\ \hline
 &  &  & \\
A  &    170     &    410 &   1420  \\
& &   & \\
A &     200  &   616 &   2133  \\
 & &   & \\ \hline
 & &   & \\
B  &   170 &   391   &  1355  \\
 & &  & \\
B  &   200 &   587   &    2034  \\ 
 & &  & \\ \hline
 & &   & \\
C  &   196 &   596    &  2066  \\
 & &   & \\
 \hline 
\end{tabular}

\end{table}

Now we would like to study the sensitivity of the  width estimates to the choice of the
LQCD pressure.  For this purpose  we will study  the model  described by 
Eq.  (\ref{pT4}). Let us assume that  the pressure of the QGP below $ T =  419.09$ MeV
is given by (\ref{pT4}) which we found for temperatures between $240.31$ and  $419.09$ MeV. Choosing $T_0$ to be $T_H$, we obtain 
\begin{align}\label{pT4T1}
\hspace*{-0.15cm}
p_{QGP}  &=   \tilde{a}_0 T^4 \ln\left[ \frac{T}{T_H} \right] - \frac{\tilde{a}_1}{3}\,  T
+  \frac{\tilde{a}_1\, T^4}{3 \,T_H^{3}}\,  \,,  
\end{align}
%
  
\noindent
since  the FWM  pressure $p^+$ (\ref{PposMn}) must vanish at $T= T_H$.
Since  the coefficient $\tilde{a}_0 \approx 0.2514 \ll 1$
is much smaller than  the Stefan-Boltzmann constant  over three $\sigma = \frac{95}{180}\pi^2 \approx 5.2 \gg 1 $ for the SU(3)$_C$ LQCD with three flavors, then
 the logarithmic term in (\ref{pT4T1}) remains   a small correction  for all temperatures
below a few tens of $T_H$ value.  Furthermore, the logarithmic term in (\ref{pT4T1}) cannot 
describe an asymptotic behavior of the LQCD pressure at large $T$.
Thus, there is only a single possibility  to match Eq.  (\ref{pT4T1}) with the LQCD 
data, namely to identify the last term in the right hand side of (\ref{pT4T1}) 
with the Boltzmann limit of the LQCD pressure at $T\gg T_H$. This condition fixes 
the value of $T_H$: 

\vspace*{-0.35cm}
\begin{align}\label{aTh}
\hspace*{-0.15cm}
\frac{\tilde{a}_1}{3 \,T_H^{3}} = \sigma \equiv   \frac{95}{180}\pi^2 \quad \Rightarrow \quad 
T_H = \left[  \frac{\tilde{a}_1}{3 \, \sigma} \right]^{\frac{1}{3}} \approx 180 ~{\rm MeV.}
\end{align}
%
  
\noindent
Matching $p^+$ and $p_{QGP}$ (\ref{pT4T1}) and expanding the logarithmic function 
at $T = T_H$, one can find the width coefficient for $T \ge c_\pm \, T_H$ as

\vspace*{-0.15cm}
\begin{align}\label{gammaII}
\gamma^2_+ &= 2 \, \beta^{-1} \biggl[ \sigma T_H T (T^2 + TT_H + T_H^2) - B(T)  \nonumber \\
&+~ \tilde{a}_0 \, T^4 \sum_{k=0}  \frac{(-1)^k}{k+1} \left(  \frac{T- T_H}{T_H} \right)^k \biggl] \,. 
\end{align}

\vspace*{-0.05cm}
\noindent
Note that in  evaluating the pressure  (\ref{pT4T1})  the coefficient $\tilde{a}_1$ 
was written to take into account  Eq.  (\ref{aTh}). 

As  in a previous case,  it is necessary that  $(T-T_H)$ is a divisor 
of the difference staying  in the square brackets in (\ref{gammaII}).
Again here  we would like to consider the simplest generalization of 
(\ref{BI})
which satisfies  the necessary condition for $T \ge c_\pm \, T_H$:

\vspace*{-0.35cm}
\begin{align}\label{BT4T1}
\hspace*{-0.15cm}
B(T) =  \sigma T_H^2  (T^2 + TT_H + T_H^2)  + \tilde{a}_0 T^{4-l} T_H^l \,,
\end{align}
%
  
\noindent
where power $l$ can be  $0; 1; 2; 3$ or $ 4$. 
Note that for $l=1$ one obtains exactly the same $T$-dependence as for  the mass density (\ref{EqXVI}) of the MIT bag model pressure and, hence,  $l=1$  is of a special interest.

Substituting (\ref{BT4T1}) into  (\ref{gammaII}) one finds

\vspace*{-0.15cm}
\begin{align}\label{gammaIII}
\gamma^2_+ &= 2 \, T T_H \biggl[ \sigma T_H  (T^2 + TT_H + T_H^2) + \tilde{a}_0 \, T^{4-l}
\frac{[T^l - T_H^l] }{T - T_H}  \nonumber \\
&-~ \frac{ \tilde{a}_0 \, T^4}{T_H}  \sum_{k=0}  \frac{(-1)^k}{k+2} \left(  \frac{T- T_H}{T_H} \right)^k \biggl] \,. 
\end{align}

\vspace*{-0.05cm}
\noindent
Assume now  that the  expressions  for the  pressure (\ref{pT4T1})  and the mass density of  bags  (\ref{BT4T1})  are valid for  $T <  c_\pm \, T_H$ as well.  These assumptions 
allow us to find the width  coefficient  in the region  $T <  c_\pm \, T_H$ 

\vspace*{-0.15cm}
\begin{align}\label{gammaIV}
\gamma^2_- &=   \frac{B(T)^2}{ 2 \left[  \sigma (T_H^3 - T^3) + \tilde{a}_0 T^3 \ln[T_H/T]  \right]} \,.
\end{align}

\vspace*{-0.05cm}
\noindent
Taking the limit $T \rightarrow 0$ in (\ref{gammaIV}) one finds the 
width coefficient at zero temperature as 

\vspace*{-0.15cm}
\begin{align}\label{gammaV}
\gamma^2_-(T = 0) &=   \frac{\left[ \sigma +  \tilde{a}_0 \, \delta_{l,4} \right]^2}{ 2   \sigma   } T_H^5 \,,
\end{align}

\vspace*{-0.05cm}
\noindent
where $\delta_{l,k}$ denotes the Kronecker symbol. The last result shows that the logarithmic term in the pressure (\ref{pT4T1}) modifies  our previous estimates  for the width coefficient  at $T= 0$  by   about 10 \% for $l = 4$ only,
whereas for $l \le 3$ and, hence, for $l=1$,  the resonance width coefficient  at $T=0$ remains unchanged. 
The corrections of the same order of magnitude are generated by $B(T)$
(\ref{BT4T1}) at $T = T_H$:
\begin{align}\label{gammaVI}
\gamma^2_+ (T=T_H) &= 2 \left[ 3\, \sigma - l\, \tilde{a}_0  \right] \, T_H^5\,.
\end{align}
Thus, the resonance  width values given in Table I remain almost the same for
the pressure  behavior as  in (\ref{pT4T1}).

\section{Asymptotic Behavior of the Regge Trajectories}

The behavior of the width of hadronic  resonances was extensively 
studied almost forty years ago  in the  Regge poles method
what was dictated by an intensive analysis of the strongly interaction dynamics in high energy hadronic 
collisions.
A lot of effort was put forward \cite{RT1,Trushevsky:77} to elucidate the asymptotical behavior of the resonance trajectories
$\alpha(S)$ for $|S| \rightarrow \infty$ ($S$ is an invariant mass square in the reaction).  Since the Regge trajectory  determines not only the mass of  resonances,
but their width as well, it would be interesting  to compare these results  with the  FWM predictions. 
Note that nowadays there is  great   interest in  the behavior of the  Regge  trajectories of higher resonances in  the  context  of the 
5-dimensional {string   theory holographically dual to QCD}  \cite{AdS} which is known as  AdS/CFT.

In our research we follow 
Ref. 
\cite{Trushevsky:77} which is based on the following most general assumptions:
(I) $\alpha(S)$ is an analytical function, having only the physical cut  from 
$S= S_0$ to $S = \infty$; (II)  $\alpha(S)$ is polynomially restricted at the whole physical sheet;  (III) there exists a finite limit of the phase trajectory at 
$S \rightarrow  \infty$.
Using these assumptions, it was possible to prove \cite{Trushevsky:77}  that  for
$S \rightarrow  \infty$ the upper bound  of  the Regge  trajectory asymptotics 
at the whole physical sheet 
is
\begin{align}\label{aUP}
\alpha_u (S) = - g^2_u \left[  - S   \right]^\nu \,,\quad {\rm with} \quad   \nu \le 1 \,,
\end{align}
where the function  $g^2_u > 0$ should increase slower than any power 
in this limit  and   its  phase must  vanish at  $|S| \rightarrow ~\infty$.

On the other hand, in Ref. \cite{Trushevsky:77}  it was also shown that, if 
in addition to (I)-(III) 
one requires  that the   transition amplitude $T(s,t)$ is 
a  polynomially  restricted function of $S$ for all nonphysical $t > t_0 > 0$,
then the real part of the Regge tragectory  does not  increase  at 
$|S| \rightarrow ~\infty$ and  the trajectory behaves as 
\begin{align}\label{aLOW}
\alpha_l (S) =  g^2_l \left[ - \left[  - S   \right]^{\frac{1}{2}} + C_l \right] \,,
\end{align}
where  $g^2_l > 0$ and $C_l$ are some constants.  Moreover,  (\ref{aLOW}) defines the lower bound for the asymptotic behavior of the Regge trajectory \cite{Trushevsky:77}. The  expression (\ref{aLOW})   is a generalization of a well known  Khuri  result \cite{RT2}. It means that for each family of hadronic  resonances  the Regge poles do not go beyond  some vertical line in the complex spin plane. 
In other words,  it means that in asymptotics $S \rightarrow + \infty$ the resonances become infinitely wide, i.e. they are moving out of the real axis of the proper angular momentum $J$ and, therefore,   there are only  a finite number of resonances in the corresponding transition amplitude.   At first glance  it seems that the huge deficit  of heavy hadronic resonances 
compared to the Hagedorn mass  spectrum 
(the second conceptual problem of  Sect. II) supports such a conclusion.
Since there are  a finite number of resonance families \cite{RegBook} it is impossible to  generate from them an exponential mass spectrum
and, hence, 
the Hagedorn mass spectrum   cannot exist 
for large resonance masses.  
Consequently, the GBM and its  followers  run  into  a deep trouble.  We, however, believe that the FWM with   negative value of the most probable bag mass $ \langle m \rangle \le  0$ can help to resolve this problem as well. 

First we note that the direct comparison of the FWM predictions with the 
Regge poles asymptotics is impossible because  the resonance mass and its 
width $\Gamma(v)$ are independent variables in the FWM. 
Nevertheless, we can relate their  average values  and compare them to the results of 
Ref. \cite{Trushevsky:77}.

{
To illustrate this statement, 
we  recall our result  on the mean Gaussian  width of the free bags averaged with respect to their volume  
by the spectrum (\ref{Rm})
(see  two paragraphs after Eq. (\ref{Rm}) for details)  
\begin{align}\label{WidthFreeBag}
\overline{\Gamma_1(v) } \approx  \Gamma_1(m/B) = \gamma
\sqrt{ \frac{m}{B} }\,. 
\end{align}
Using the formalism of  \cite{Trushevsky:77}, it  can  be shown that 
at zero temperature
the free QGP bags of mass $m$ and mean resonance  width $\alpha\, \overline{\Gamma_1(v) } |_{T=0} \approx \alpha\,\gamma_0
\sqrt{ \frac{m}{B_0} }$  
precisely correspond  to  the following Regge trajectory 
\begin{align}\label{alphaHT}
&  \hspace*{-0.1cm}
\alpha_r  (S) =  g^2_r [S + a_r (- S)^\frac{3}{4} ] \quad  {\rm with} \quad {a_r =const  < 0}\,. 
\end{align}
Indeed, substituting $S = |S| e^{i \, \phi_r}  $ into (\ref{alphaHT}), then  expanding  the second term on the right hand side of   (\ref{alphaHT}) and requiring ${\rm Im} \left[ \alpha_r (S) \right] = 0$, one finds the phase of  
physical trajectory (one of four roots of one fourth power in  (\ref{alphaHT}))
\begin{align}\label{phiS}
&  
\phi_r  (S) \rightarrow  \frac{a_r \sin \frac{3}{4}\pi  }{|S|^\frac{1}{4} }  \rightarrow 
0^-\,,  
\end{align}
which is vanishing in the correct quadrant of the complex $S$-plane. 
Considering the complex energy plane $E = \sqrt{S} \equiv M_r - i \frac{\Gamma_r}{2}$, 
one can  determine the mass  $M_r$ and the width $\Gamma_r$  
\begin{align}\label{mgamI}
&  
\hspace*{-0.290cm}
M_r \approx |S|^\frac{1}{2} \, {\rm and } \,\, \Gamma_r \approx - |S|^\frac{1}{2}
\phi_r  (S) =  \frac{|a_r| |S|^\frac{1}{4}   }{  \sqrt{2} }  =  \frac{|a_r| M_r^\frac{1}{2}   }{  \sqrt{2} },
\end{align}
of a resonance belonging to the trajectory  (\ref{alphaHT}).

Comparing the mass dependence of the width in (\ref{mgamI}) with the mean width of  free QGP bags 
(\ref{WidthFreeBag}) taken at $T=0$, it is natural  to identify them
\begin{align}\label{aIpfree}
&
a_r^{free} \approx - \alpha \,\gamma_0 \, \sqrt{\frac{2}{B_0}} = - 4  \, \gamma_0 \, \sqrt{\frac{\ln 2}{B_0}}  \,, 
\end{align}  
and to deduce  that   the free QGP bags belong to the Regge trajectory (\ref{alphaHT}).
Such a conclusion is in line both with the well established results on the linear  Regge trajectories of hadronic resonances \cite{RegBook} and
with  theoretical expectations  of  the dual resonance model \cite{DualRM}, the open string model 
\cite{Shuryak:sQGP,StringW}, the closed string model \cite{Shuryak:sQGP} and  the   AdS/CFT \cite{AdS}.
Such a property of the FWM also  gives   a very strong argument in favor of  both  the 
volume dependent width $\Gamma (v) = \Gamma_1(v)$ and  
the corresponding mass-volume spectrum of heavy bags 
(\ref{Rfwm}).
}

Next we consider the second  way of averaging the mass-volume spectrum 
with respect to  the resonance  mass
\begin{align}\label{mmeanI}
& \hspace*{-.8cm}
\overline{m} (v)  ~ \equiv ~  \frac{   \int\limits_{M_0}^{\infty}\hspace*{-0.0cm} dm 
\int  \frac{d^3k}{(2\pi)^3}  \,\rho(m,v) ~ m ~ e^{- \frac{\sqrt{k^2 + m^2} }{T}} 
}{
 \int\limits_{M_0}^{\infty}\hspace*{-0.cm} dm
\int  \frac{d^3k}{(2\pi)^3}  \,\rho(m,v) ~e^{- \frac{\sqrt{k^2 + m^2} }{T}} 
 }
\, ,
\end{align}
which is technically simpler than averaging with respect to the resonance volume,
but we will make  the necessary  comments  on the other way of  averaging  in the  appropriate places.

Using the results of Sect. IV one can find the mean mass (\ref{mmeanI})
for $T \ge  c_\pm\, T_H$ (or for $ \langle m \rangle \ge 0 $ ) to be equal to 
the most probable mass of  bag from which one determines the resonance width:
\begin{align}\label{mmassII}
&  \hspace*{-.25cm}
\overline{m} (v)   \approx  \langle m \rangle  \quad {\rm and} \\
\label{mgammaII}
&  \hspace*{-.25cm}
\Gamma_R (v) \approx  2 \sqrt{2 \ln2 }\, \Gamma_1 \hspace*{-.1cm} \left[ 
\frac{ \langle m \rangle  }{\textstyle B + \gamma^2 \beta}  \right] 
\hspace*{-.1cm}  = 2 \,\gamma \sqrt{ 
\frac{2 \ln 2 \,\langle m \rangle }{B + \gamma^2 \beta}  }  \,.
\end{align}
The last two equations lead to a vanishing  ratio 
$\frac{\Gamma_R}{\langle m \rangle} \sim  \langle m \rangle^{-\frac{1}{2}}$ in the limit $\langle m \rangle \rightarrow \infty$.
Comparing (\ref{mmassII}) and (\ref{mgammaII}) with the mass and width  (\ref{mgamI}) of the 
Regge trajectory (\ref{alphaHT}) and applying absolutely the same logic which we used for the 
free QGP bags, we conclude 
that the location of the FWM heavy bags in the complex energy plane 
is identical to that one of resonances belonging to the trajectory 
(\ref{alphaHT}) with 
\begin{align}\label{mgamII}
&  
\langle m \rangle \approx |S|^\frac{1}{2} \quad {\rm and } \quad 
a_r \approx  - 4 \gamma \, \sqrt{  \frac{\ln 2   }{  B + \gamma^2 \beta }  }
\,.
\end{align}
The  most remarkable output of such a conclusion  is that   the medium dependent FWM  mass and  width of the extended QGP bags obey the upper  bound  for  the Regge
trajectory asymptotic  behavior   obtained for point-like hadrons 
\cite{Trushevsky:77}!

It is also  interesting that the resonance  width formula (\ref{mgammaII}) is generated by the most 
probable volume  
\begin{align}\label{vem}
v_E (m) \approx \frac{m}{\sqrt{B^2 + 2 \gamma^2 s^*} } = \frac{m}{B + \gamma^2 \beta}
\,.
\end{align}
of heavy resonances of mass $m \gg M_0$ that are described by the continuous  spectrum $F_Q (s,T)$   (\ref{FsHQ}). 
This result can be  easily found by maximizing   the exponential in $F_Q (s,T)$   
with respect to resonance volume $v$ at 
fixed mass $m$ and  by recalling that at high temperatures 
the rightmost singularity  of the isobaric partition  (\ref{Zs}) is defined by the pressure 
(\ref{PposMn}) as  $s^* = \frac{p^+}{T}$.

The extracted  values of the  resonance width coefficient along with the relation  (\ref{BI}) for  $B (T)$ allow us to estimate $a_r$  as 
\begin{align}\label{aR}
& 
a_r \approx  - 4  \, \sqrt{   \frac{2\,    T\, T_H \,  }{ 2\, T - T_H }   \ln 2 } \,. 
\end{align}
This expression shows  that for $T \rightarrow T_H/2  + 0$  the asymptotic behavior 
(\ref{alphaHT}) breaks down since the resonance width  diverges  at fixed $|S|$. On the other hand  from (\ref{aR}) it follows that  $a^2_r (T= T_H)  \approx  22.18 \, T_H $
and  $a^2_r (T\gg T_H)  \approx  11.09 \, T_H $. In other words,  for a typical value of 
the Hagedorn  temperature  $T_H \approx 190 $ MeV (see a discussion after Eq. (19)) 
(\ref{aR}) gives a reasonable range of the invariant mass  $|S|^\frac{1}{2} \gg a^2_r (T= T_H)  \approx 4.21 $ GeV  and  $|S|^\frac{1}{2} \gg a^2_r (T \gg T_H)  \approx 2.1 $ GeV
for which  Eq.   (\ref{alphaHT}) is true.

Now we can find the spin of the FWM  resonances 
\begin{align}\label{Jm}
& 
J  = {\rm Re} \, \alpha_r (\langle m \rangle^2) \, \, \approx   g_r^2  \,
 \langle m \rangle \left[ \langle m \rangle - \frac{a_r^2}{4} \right] \,,
\end{align}
which  has  a typical Regge behavior up to a small correction.
Such a property can also be  obtained   within  the dual resonance model \cite{DualRM},
within the models of  open  \cite{Shuryak:sQGP}  and  closed  string \cite{Shuryak:sQGP,StringW}, and within 
 AdS/CFT \cite{AdS}.
These models  support our result (\ref{Jm}) and justify  it.  Note, however, that 
in addition to the spin value  the FWM determines the width of hadronic  resonances. 
The latter  allows us to predict   the  ratio 
of widths of two resonances having spins $J_2$ and $J_1$
and appearing at the same temperature $T$  to be  as follows
\begin{align}\label{g2g1}
& 
\frac{
\Gamma_R \biggl[ \frac{\langle m \rangle\bigl|_{J_2}}{ (B + \gamma^2 \beta) } \biggl]}{
\Gamma_R \biggl[ \frac{\langle m \rangle\bigl|_{J_1}}{ (B + \gamma^2 \beta)} \biggl]
}
\approx \frac{ \sqrt{v \bigl|_{J_2} }  }{ 
\sqrt{v \bigl|_{J_1}  } }
\approx \frac{ \sqrt{\langle m \rangle\bigl|_{J_2} }  }{ 
\sqrt{\langle m \rangle\bigl|_{J_1}  } } \approx   \left[  \frac{J_2}{J_1} \right]^\frac{1}{4}\,,
\end{align}
which, perhaps, can  be tested at LHC.

Now we turn to the analysis of the low temperature regime, i.e. to 
$T \le c_\pm T_H$.  Using previously obtained results from  (\ref{mmeanI}) one finds 
\begin{align}\label{mmassIII}
& 
\overline{m} (v)  ~ \approx ~  M_0 \,,
\end{align}
i.e.
the mean mass is volume independent. Taking the limit $v \rightarrow \infty$, we get the ratio $\frac{\Gamma(v)}{\overline{m} (v)} \rightarrow  \infty $ which closely resembles  the case of  the  lower bound  of the Regge trajectory asymptotics  (\ref{aLOW}). 
Similarly to the analysis of  high temperature regime, from (\ref{aLOW}) one can find  the trajectory  phase and then  the resonance  mass $M_r$ and 
its width $\Gamma_r$ 
\begin{align}\label{phiSL}
&  
\phi_r  (S) \rightarrow - \pi + \frac{2 |C_l |  |\sin (\arg C_l) |  }{|S|^\frac{1}{2} }  \,,\\
\label{mgamL}
& 
M_r \approx   |C_l |  |\sin (\arg C_l) | \quad {\rm and} \quad
\Gamma_r \approx  2 |S|^\frac{1}{2} \,. 
\end{align}
Again comparing the  averaged  masses  and width of FWM resonances 
with  their counterparts in  (\ref{mgamL}), we  find  similar behavior in the 
limit of large  width of resonances.  Therefore, we 
conclude that 
at low temperatures the FWM obeys the lower bound of the Regge trajectory asymptotics of  \cite{Trushevsky:77}.

The other way of averaging, i.e. with respect to the resonance volume,  
in the leading order   gives  the most probable resonance  volume of the continuous  spectrum $F_Q (s,T)$   (\ref{FsHQ}) defined by 
 the left equation  (\ref{vem}) again.  Substituting in it  the corresponding rightmost singularity 
$s^*= \frac{p^-}{T}$ and using  (\ref{pnegM2}) for $p^-$,  one finds $v_E(m) \rightarrow \infty$ which leads to an infinite value of the most probable resonance width defined in this way. 
Note  that such a result is supported by  the high temperature mean width behavior  if   $T \rightarrow T_H/2 + 0$.
As one can see from (\ref{aR}) and (\ref{mgamI}),   in the latter case   the trajectory  (\ref{alphaHT})   also demonstrates 
a very large width compared to a finite resonance  mass.

A  more refined analysis of the most probable volume (\ref{vem}) shows that the second derivative of the exponential in $F_Q (s,T)$ with respect to the resonance volume vanishes at large resonance masses $m$ and, hence,   one needs  to account for   even  weaker dependences on resonance volume $v$ and  to  inspect   higher   order derivatives with respect to $v$, but this task is out of the scope of present work 
and we leave it for future 
investigation.

Thus, these estimates  demonstrate  that at any temperature  the FWM QGP bags can be regarded as the medium induced Reggeons which at $T \le c_\pm T_H$ 
(i.e. for  $\langle m \rangle \le 0$) belong to the Regge trajectory   (\ref{aLOW}) and otherwise they are described by the  trajectory (\ref{alphaHT}).
Of course, both of the trajectories (\ref{aLOW}) and  (\ref{alphaHT})
are valid in the asymptotic $|S| \rightarrow \infty$, but the most remarkable  fact is that,
to our knowledge, the FWM gives us the first example of a model which reproduces both  of these trajectories and, thus, obeys both bounds of the Regge asymptotics.  Moreover, since the FWM contains the Hagedorn-like mass spectrum at any temperature, the Subthreshold Suppression of QGP bags removes the contradiction between the Hagedorn  ideas on the exponential mass spectrum of hadrons   and the Regge
poles method in the low temperature domain! Furthermore, the FWM opens a possibility to apply the Regge poles  method to a variety of processes in a strongly interacting  matter  and account, at least partly,  for 
some of the medium effects.



\section{Conclusions and Perspectives}


Here we present  the novel statistical approach to study  the QGP bags with  medium dependent width. 
We argue   that the volume dependent width of the QGP bags $\Gamma (v) = \gamma\, v^\frac{1}{2}$ leads to the Hagedorn mass spectrum of heavy bags.  Such  behavior of a width  allows us to explain a huge  deficit of heavy hadronic resonances in the experimental mass spectrum. 
The key point of our treatment   is the presence of Gaussian  attenuation of bag mass. 
Perhaps,  the nonlocal field theoretical models may shed light on the origin
of the Gaussian mass attenuation. 

Under  plausible  assumptions we derive  the general  expression  for  the bag pressure $p^+$ which accounts for the effect of finite  width in the EoS.
We argue  that the obtained spectrum itself
cannot  explain the absence of directly observable QGP bags and strangelets  in the high energy nuclear 
and elementary particle collisions. Then we demonstrate  the  possibility to ``hide''  the heavy  QGP bags for $T \le c_\pm  T_H$  by their {\it subthreshold suppression}. The latter occurs due to the fact 
that at low temperatures the most probable mass of heavy bags $\langle m \rangle$  becomes negative and, hence, is below the lower cut-off  $M_0$ of
the continuous mass spectrum. 
Consequently,  only the lightest  bags of mass about $M_0$ and of  smallest volume $V_0$
may contribute into the resulting spectrum, but such QGP bags will be indistinguishable  
from the low-lying  hadronic resonances with the short life-time. 
We show  how  the FWD can reproduce a few  EoS of the  QGP  and discuss the corresponding restrictions. 

We analyze   the recent LQCD data for the trace anomaly  and extract 
the linear $T$-dependent term for the LQCD pressure  and higher order corrections to it. 
This linear $T$-dependent term in the LQCD pressure is naturally associated with the FWM 
pressure at low temperatures $p^-$. Using such a dependence we estimate 
the volume dependent width under plausible assumptions and 
find it almost insensitive  to the number of color and flavor degrees of freedom of the LQCD data. 
These 
estimates clearly demonstrate   that  such short  living QGP bags cannot be established  experimentally. 
We believe that our  finding  introduces the new time scale
into the high energy nuclear and elementary particles collisions 
and requires some modifications of the present picture of the 
collision process and its subsequent stages.

With the help of formalism of \cite{Trushevsky:77}  we   show  that  the average mass and width of  heavy or large   free QGP bags  belong to the linear Regge trajectory (\ref{alphaHT}).
Similarly, we find that
at  hight temperatures the average mass and width of the QGP bags  behave in accordance with  the upper bound of the Regge trajectory asymptotics (\ref{aUP}) (linear trajectory), whereas at low temperatures they obey  the lower bound 
of   the Regge trajectory asymptotics (\ref{aLOW}) (square root trajectory). Since the model explicitly contains the Hagedorn spectrum,  it  removes an existing contradiction between the finite number of 
hadronic Regge  families
and the Hagedorn idea of the exponentially growing  mass spectrum of hadronic bags.  
Such a result creates a new look onto the large and/or heavy   QGP bags as the medium induced  Reggeons 
and  opens a principal  possibility to apply all the strength of  the Regge poles  method to a variety of processes in a strongly interacting  media and to account, at least partly,  for  some of the medium effects.

Besides these general results  the FWM  allows us to make some  predictions which  can be  soon tested experimentally.
Thus, the  relation  between the maximal  spin of the bag and the most probable 
mass (\ref{Jm}), or the dependence between the maximal spin  and the mean volume  (\ref{g2g1}) can be, perhaps, tested at LHC 
CERN during the hadronic collisions runs.  It is also probable that  the switch 
between  the Regge trajectories  (\ref{aLOW}) and (\ref{alphaHT}) can be verified 
at the FAIR  GSI and NICA JINR energy range.  It is clear that the lower bound 
trajectory  (\ref{aLOW}) cannot be established  in a laboratory, but it seems reasonable to expect that  the experimentalists  will be able to  find how  the parameters of the upper bound trajectory  (\ref{alphaHT})  are  changing with the colliding energy. 
The found parameters of bag trajectories  in combination with the HBT analysis of their  volume may help us to determine the resonance  width coefficient $\gamma$ or even  the  volume dependence of the resonance width itself.  

The generalization of all of the above  results  to  non-zero baryonic densities can be made  straightforwardly  by assuming 
the dependence of the model functions $B$ and $\Gamma_1 (v)$  on the baryonic  chemical potential. 
The more detailed quantitative  experimental consequences of the FWM  will be published elsewhere \cite{Bugaev:0809}.

{\bf Acknowledgments.} We are thankful  to P. Braun-Munzinger, 
P. J.  Levai,  E. S. Martynov,  D. H.  Rischke,
J. Stachel, O. V. Teryaev and D. Voskresensky  for fruitful discussions and important comments.
One of us, K.A.B., thanks the Laboratory for Theoretical Physics of JINR, Dubna, where  part of  this work was completed, 
 for  warm hospitality.  
The research made in this paper  
was supported in part   by the Program ``Fundamental Properties of Physical Systems 
under Extreme Conditions''  of the Bureau of the Section of Physics and Astronomy  of
the National Academy of Science of Ukraine.







\end{document}